\documentstyle[aps,prl,twocolumn,psfig,floats]{revtex}

\begin{document}

\draft

\title{Adiabatic passage by light-induced potentials in molecules}

\author{B. M.~Garraway\cite{endnote}}

\address{Optics Section, Blackett Laboratory, Imperial College,
Prince Consort Road, London, SW7~2BZ, United Kingdom}

\author{K.-A.~Suominen}

\address{Helsinki Institute of Physics, PL 9,
FIN-00014 Helsingin yliopisto, Finland}

%\date{\today}

\maketitle

\narrowtext

\begin{abstract}
We present the APLIP process (Adiabatic Passage by Light Induced Potentials)
for the adiabatic transfer of a wave packet from one molecular potential to
the displaced ground vibrational state of another. The process uses an
intermediate state, which is only slightly populated, and a counterintuitive
sequence of light pulses to couple the three molecular states. APLIP shares
many features with STIRAP (stimulated Raman adiabatic passage), such as high 
efficiency and insensitivity to pulse parameters. However, in APLIP there is 
no two-photon resonance, and the main mechanism for the transport of the wave 
packet is a light-induced potential. The APLIP process appears to violate the 
Franck-Condon principle, because of the displacement of the wave packet, but 
does in fact take place on timescales which are at least a little longer than 
a vibrational timescale.
\end{abstract}


Femtosecond pulses have recently opened the possibility to create, observe and
control the internal dynamics of molecules~\cite{reviews,rpp}. Typically one
has studied the pump-probe situation, where a molecular wave packet, i.e., a
coherent superposition of vibrational states, has been created by the first
pulse, and a second pulse probes the subsequent evolution of the wave packet.
Alternatively one can observe the dissociation process directly. In more subtle
cases, such as the pump and dump schemes, the final result can be some special
molecular bound state instead of dissociation. Such manipulations provide new
understanding of molecular dynamics and chemical reactions. Also, they present
intriguing demonstrations of wave packet dynamics and time-dependent quantum
mechanics in general.

\begin{figure}[tbh]
\centerline{
\psfig{width=90mm,file=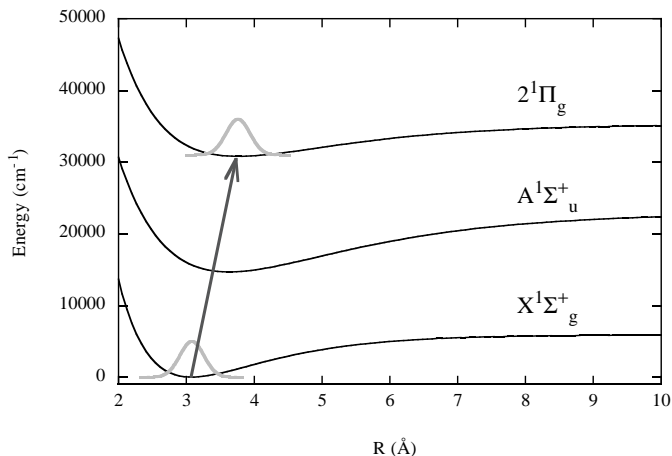}
}
\caption[]{
The three Na$_2$ potential energy surfaces used in our calculations: the
X$^1\Sigma_g^+$ ground state, the A$^1\Sigma_u^+$ state as the first
excited state, and the 2$^1\Pi_g$ as the third state. The diagonal sloping
arrow indicates the overall effect of the two laser pulses used.
\label{fig1}}
\end{figure}

The purpose of this Letter is to demonstrate (theoretically) a mechanism for
the transfer of a stationary ground state vibrational wave packet to a {\em
stationary and displaced}  excited state wave packet through an intermediate
state which  is not significantly populated during the process. The overall 
effect is symbolically represented by the diagonal arrow in Fig.~\ref{fig1}. 
Of course, we can change from a wave packet picture of the process to a picture
in terms of the vibrational levels, in which case Fig.~\ref{fig1} illustrates 
a process where the population of the $\nu=0$ vibrational level of the ground 
state is directly transferred to the $\nu''=0$ vibrational level of the second 
excited state. The overall effect appears as a violation of the Franck-Condon 
principle, which can be simply stated as saying that there should only be 
vertical transitions between vibrational states in a molecule, at least, over 
short times. Thus the diagonal transition seen in Fig.~\ref{fig1} should not be
allowed. It is further inhibited by the fact that the overlap between the
initial and final wave functions is very small (the Franck-Condon overlap). 
This is because the initial wave packet is displaced over a distance of about 
seven times its width for the example of Fig.~\ref{fig1}.  Of course, there is 
no real violation of the Franck-Condon principle; we manipulate the molecular 
states on timescales close to, but longer than the vibrational timescales.

To illustrate the process we have chosen the sodium dimer, a molecule
which has already been subjected to much study in the field of wave
packet dynamics, and which opens the prospect of an experiment to test
the ideas in this Letter. Following Refs.~\cite{baumert} we have chosen
our three states so that the ground state is the X$^1\Sigma_g^+$ of Na$_2$,
the first excited state is the A$^1\Sigma_u^+$ state, and the second
excited state is the 2$^1\Pi_g$ state. Data for the molecular potentials
has been gathered from Refs.~\cite{baumert,potsources} and this data is used in
a numerical calculation of the dynamics of the wave packet during the
interaction of the system with two laser pulses.

In terms of the electronic potentials the Hamiltonian for the vibrational
motion of the molecule is
\begin{equation}
   H = -\frac{\hbar^2}{2m}  \frac{ \partial^2 }{ \partial R^2 }\
       {\cal I} + {\cal U}(R,t)
   \label{molh}
\end{equation}
where $R$ is the internuclear separation, $m$ is the reduced mass of the 
molecule, and the electronic potentials and couplings are given by
\begin{equation}
   {\cal U}(R,t) = \left[
   \begin{array}{ccc}
      U_X(R)+\hbar\delta_1 & \hbar\Omega_1(t)  & 0 \\
      \hbar\Omega_1(t)  & U_A(R) & \hbar\Omega_2(t) \\
      0 & \hbar\Omega_2(t)   & U_\Pi(R)+\hbar\delta_2
   \end{array} \right].
   \label{U}
\end{equation}
Here $U_X(R)$, $U_A(R)$, and $U_\Pi(R)$ are the three potentials, $\delta_1$
and $\delta_2$ are the detunings of the two pulses from the lowest points of
the potentials, and $\Omega_1(t)=\mu_{XA} E_1(t)/\hbar$,
$\Omega_2(t)=\mu_{A\Pi} E_2(t)/\hbar$ are the two Rabi frequencies.
We have assumed for simplicity that the two dipole moments are independent of
$R$ and we solve the time-dependent Schr\"odinger equation with Hamiltonian
(\ref{molh}) by a numerical method (see, e.g. Ref.\
\cite{rpp}).

\begin{figure}[tbh]
\vspace*{-2.5cm}
\noindent\centerline{
\psfig{width=90mm,file=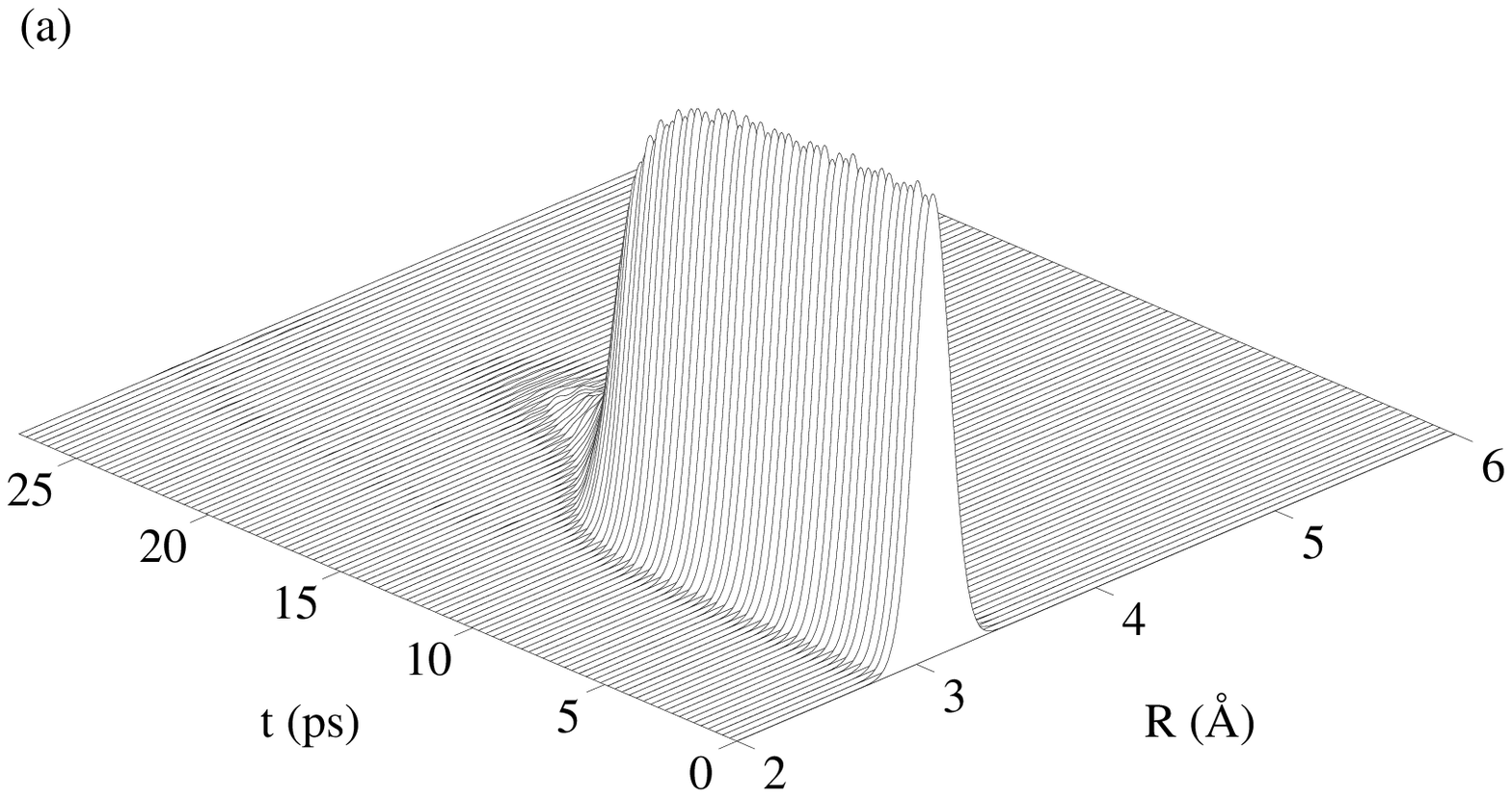}
}
\vspace*{-1cm}

\noindent\centerline{
\psfig{width=90mm,file=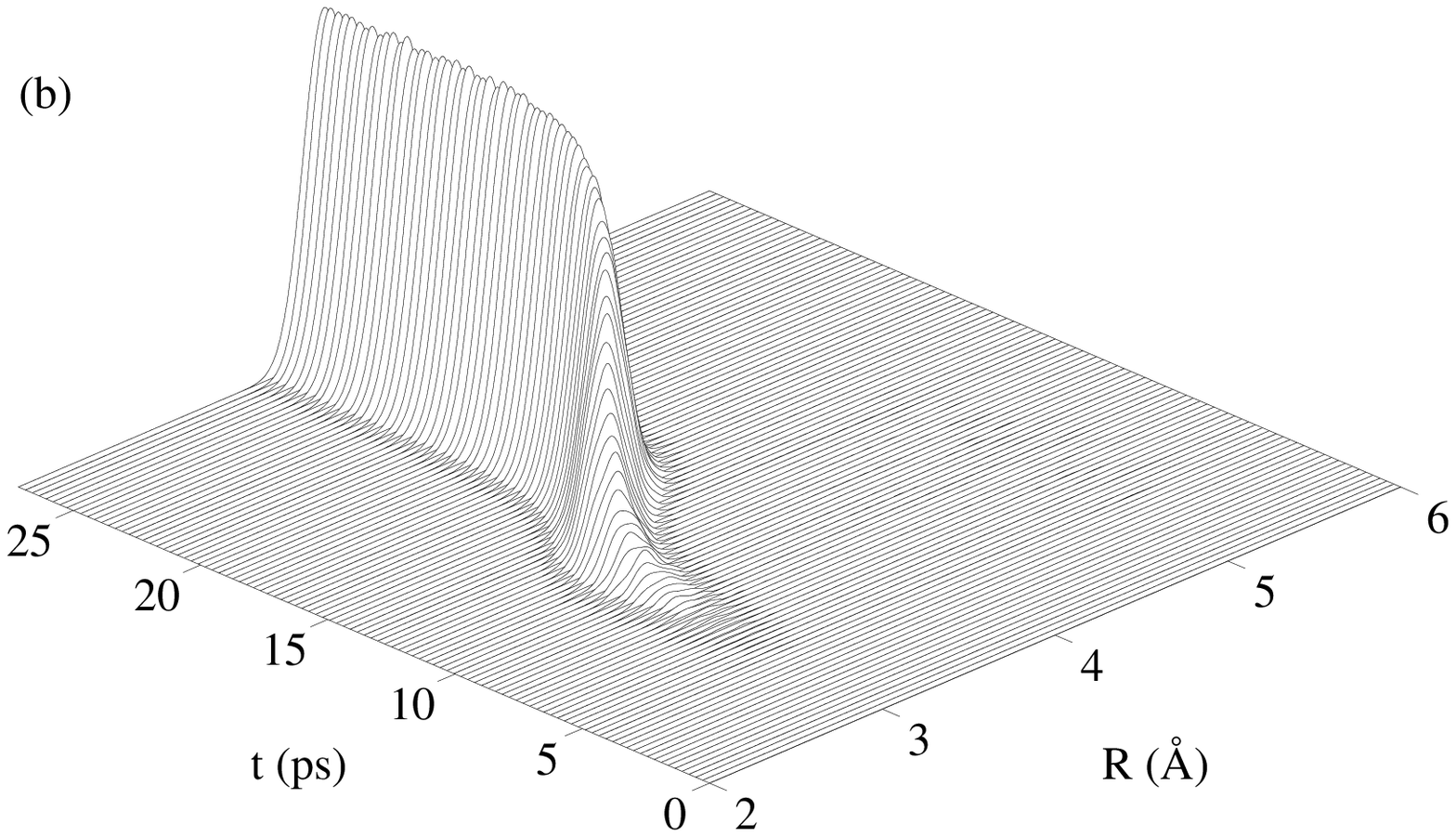}
}
\caption[]{
The wave packet dynamics on (a) the X$^1\Sigma_g^+$ ground state and
(b) the target state  2$^1\Pi_g$. The wave packet motion has been determined
from a fully quantum mechanical calculation. In (a) the ground state
wave packet disappears when the two pulses arrive at t=10.8 ps and 16.3 ps.
In (b) we see the slow and steady appearence of the wave packet in the
2$^1\Pi_g$ state. Note the steady displacement of the wave packet as it arrives
adiabatically to the bottom of the lowest vibrational state of the 2$^1\Pi_g$
state. Both pulses are red detuned from the bottom of the potential energy
surface by 2200 cm$^{-1}$ and have a width of 5.42 fs.
\label{fig2}}
\end{figure}

Figure~\ref{fig2} shows an example of the wave packet dynamics following
the coupling of two pulses between each pair of molecular states.
In Fig.~\ref{fig2}(a) we see at $t=0$ the initial ground state wave packet, 
located at the equilibrium position of 3.08~\AA, in the X$^1\Sigma_g^+$ 
potential. When the two Gaussian laser pulses (with peak intensities of 3 and 
6 TW/cm$^2$) act on the molecule we see the disappearance of the wave packet
from the X$^1\Sigma_g^+$ state, while it is also displaced
to the right (to larger distances). As the wave packet disappears from the
X$^1\Sigma_g^+$ state it appears on the excited 2$^1\Pi_g$
state [see Fig.~\ref{fig2}(b)], still being displaced to longer bond
lengths as it appears. When the laser pulses have completed, the wave packet
is left in the 2$^1\Pi_g$ state without any vibrational excitation (there is
no motion of the wave packet).

During the process seen in Fig.~\ref{fig2} the X$^1\Sigma_g^+$ and 2$^1\Pi_g$
states exchange their population while the population of the A$^1\Sigma_u^+$
state remains very low at all times. The process is nearly 100\% efficient in
transferring population from the X$^1\Sigma_g^+$ state to the 2$^1\Pi_g$ state.
This efficiency remains high over a wide range of pulse parameters.

The process we have described uses a counterintuitive pulse sequence: the
pulse nearly resonant with the  X$^1\Sigma_g^+ \rightarrow$ A$^1\Sigma_u^+$
transition is applied {\it after} the pulse nearly resonant with the
A$^1\Sigma_u^+ \rightarrow 2^1\Pi_g$ transition. However, the process is not
the same as the conventional STIRAP~\cite{stirapmany,stirapexp,shore95}
(Stimulated Raman Adiabatic Passage) process, already seen in molecular
systems~\cite{stirapexp}, for both trivial and fundamental reasons. The most
trivial difference with the existing experiments is the linkage pattern; the
ladder system we consider (Fig.~\ref{fig1})  has a different linkage pattern
from the Raman type {\sf $\Lambda$} system after which STIRAP is named.
In the absence of spontaneous emission~\cite{spont}, these different linkage
patterns do not affect the dynamics in the case of atomic systems
\cite{shore95}.

Because we utilize a ladder system it makes sense to consider a transition from
X$^1\Sigma_g^+$ $(\nu=0)$ to 2$^1\Pi_g$ $(\nu''=0)$, i.e., from the ground
vibrational state of the lowest electronic state to the ground vibrational
state of the highest electronic state in our three-level system. In a {\sf
$\Lambda$}-type Raman scheme this would not make sense as STIRAP is then used 
to create an excited vibrational state ($\nu\ne 0$) from the ground state
($\nu=0$) within the same electronic state of the molecule.

Conventional STIRAP utilizes a two-photon resonance condition. For example, a
suitable Hamiltonian for STIRAP in an atomic ladder system is
\begin{equation}
   H_a = \left[ \begin{array}{ccc}
           0             & \hbar\Omega_1(t) & 0 \\
           \hbar\Omega_1(t)   & \hbar\Delta      & \hbar\Omega_2(t) \\
           0             & \hbar\Omega_2(t) & 0
         \end{array} \right],
         \label{atomich}
\end{equation}
where $\Delta$ is the two-photon resonant laser-atom detuning, and $\Omega_1$
and $\Omega_2$ are the Rabi frequencies of the pump pulse and Stokes pulse.
If the pump and Stokes pulses change sufficiently slowly, we can consider
the process to be adiabatic. Then we can utilize the instantaneous eigenstate
\begin{equation}
   \psi_{z}(t) =
         \frac{1}{\sqrt{\Omega_1^2(t) + \Omega_2^2(t) }}
         [\Omega_2(t), 0, -\Omega_1(t)]^{\top} \label{psid}
\end{equation}
to achieve the transfer directly from state~1 to state~3. We note that the state
$\psi_z$ is for all $t$ an exact eigenstate of $H_a$, Eq.~(\ref{atomich}),
with eigenvalue zero (sometimes called a `dark state'~\cite{Arimondo}). If the
pulses are in the counterintuitive order, i.e., $\Omega_2(t)$ reaches its peak
before $\Omega_1(t)$, the eigenstate~(\ref{psid}) matches the initial state of
the system (state~1). Since, for long pulses, the system state adiabatically
follows the state $\psi_z$, the occupation probability is transported from
state~1 to state~3. Because state~2 is not involved in the eigenstate $\psi_z$,
it is not populated during the pulse sequence.

The situation for the molecule is rather different because we have
an extra degree of freedom: the molecular co-ordinate which we
denote by $R$. With the Hamiltonian now given by Eq.~(\ref{molh}),
and the spatially varying potentials~(\ref{U}), it is clear that
it is impossible in this molecular case to have the two-photon resonance
condition used in the atomic case (except at isolated points).
Thus there is no zero eigenstate in the molecular situation.

At this point it could be argued that rather than viewing
the Hamiltonian~(\ref{molh}) in the position basis, we should utilize a
vibrational basis so that we could recover a version of the atomic STIRAP
process seen with the Hamiltonian~(\ref{atomich}). However, whilst the
vibrational picture and the spatial picture are entirely equivalent, we
believe that the key to understanding the phenomenon in Fig.~\ref{fig2} is not
the vibrational basis but a spatial picture. This brings us to a fundamental
difference between STIRAP and the phenomena in this Letter. If we had a STIRAP
process taking the system from the X$^1\Sigma_g^+$ $(\nu=0)$ vibrational state
to the 2$^1\Pi_g$ $(\nu''=0)$ vibrational state we would only see the
disappearance of the wave packet in Fig.~\ref{fig2}(a) and its
reappearance in Fig.~\ref{fig2}(b) without the smooth positional shifting
of the wave packet. The reason is that the X$^1\Sigma_g^+$ $(\nu=0)$
vibrational state wave function will be approximately the ground
vibrational wave function of a harmonic oscillator, and
any positional movement of the wave packet must be due to the excitation
of other vibrational states. We can say the same thing about the
2$^1\Pi_g$ $(\nu''=0)$ vibrational state, i.e., that if only the $(\nu''=0)$
were involved there would be no shifting of the wave packet as seen in
Fig.~\ref{fig2}(b). So the process of Fig.~\ref{fig2} is not direct
STIRAP transfer between  X$^1\Sigma_g^+$  and  2$^1\Pi_g$.

\begin{figure}[tbh]
\centerline{
\psfig{width=82mm,file=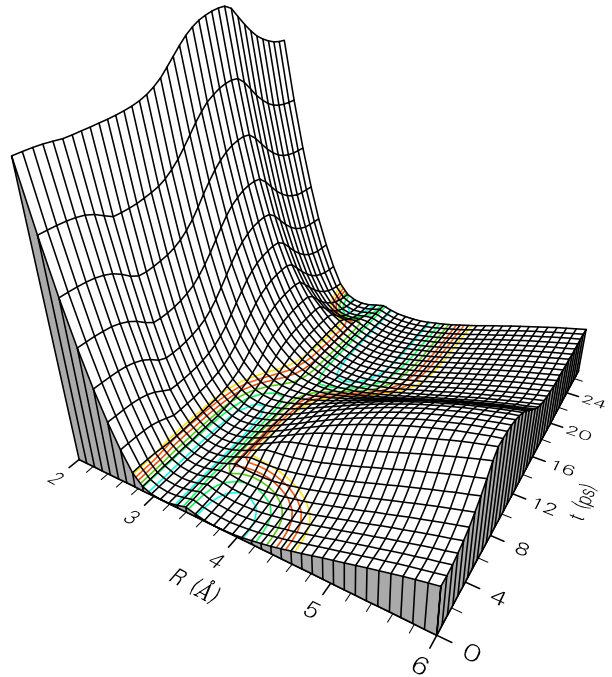}
}
\caption[]{
Space and time dependence of the light-induced potential responsible
for the transportation of the wave packet in Fig.\ \ref{fig2}.
\label{fig4}}
\end{figure}

Our explanation for the transfer of the wave packet in the manner seen in
Fig.~\ref{fig2} relies on {\it light-induced potentials}~\cite{lips}. For wave
packets that travel sufficiently slowly through systems of coupled potential
surfaces the nature of the field-dressed potential energy surfaces becomes more
important than the bare (i.e., not coupled by light) energy surfaces. This
means, for example, that a laser induced crossing becomes an avoided crossing
with an energy gap which increases with the intensity of the laser. The energy
gap can become large enough to allow the passage of a wave packet which would
not otherwise penetrate the crossing; this leads to bond softening~\cite{soft}.
The eigenvalues of~(\ref{U}) determine the light-induced potentials and in
Fig.~\ref{fig4} we show the most relevant one as a function of space and time.
The most striking feature is the kinked channel which is responsible for
guiding the wave packet from one position to another.

At $t=0$ the light-induced potential of Fig.~\ref{fig4} is comprised of the
X$^1\Sigma_g^+$ potential on the left hand side of the picture, and of the
2$^1\Pi_g$ potential on the right hand side of the picture. In effect there is
a very small avoided crossing of these two potentials near $R=3.4$~\AA. The
crossing is small because at $t=0$ the two laser fields are very weak. As
a result there are two wells in the  field-dressed state at $t=0$, one
belonging to the X$^1\Sigma_g^+$ state, where the initial wave packet resides,
and the other belonging to the 2$^1\Pi_g$ state, which is where we aim to
transfer the wave packet. Because of the red detuning of the two pulses, the
field-dressed state corresponding to the intermediate A$^1\Sigma_u^+$ state (at
$t=0$) lies well below the field-dressed potential of Fig.~\ref{fig4}. However,
as the pulse resonant with the A$^1\Sigma_u^+ \rightarrow$ 2$^1\Pi_g$
transition turns on there is a repulsion between the A-type (A$^1\Sigma_u^+$ at
$t=0$) field-dressed state and the rhs of the field-dressed state in
Fig.~\ref{fig4} (which is $\Pi$-like). This repulsion results in the
disappearance of the rhs channel in Fig.~\ref{fig4}. The repulsion also moves
the A-$\Pi$ avoided crossing to larger internuclear separations.

When the second pulse is turning on, the repulsion between the X part of the 
state in Fig.~\ref{fig4} and the lower A state pushes the main channel upwards 
in energy and also displaces it to the right because the right hand part of 
the state in Fig.~\ref{fig4} has more of the $\Pi$ character, and is not 
repelled from the A state by the second pulse. As the first pulse dies away 
the transfer of the wave packet completes with the approach of an avoided 
crossing from {\it small }internuclear separations. Eventually this avoided 
crossing becomes the same X-$\Pi$ avoided crossing seen near $t=0$, so we again
have two wells in the potential.

It can be clearly seen from Fig.~\ref{fig4} that we need to have a
counterintuitive sequence of pulses. Having an intuitive sequence would be
like starting from the rear of Fig.~\ref{fig4} in the left hand well. But the
left hand well is a dead end. Only the right hand well lies in the channel
connecting the wave packet through the pulse sequence. Because the
process is carried out slowly (adiabatically), the original wave packet can not
only be transported from one space position to another but also can change its
shape at the same time. For example, if the $\Pi$ state had a much narrower
potential (high vibrational frequency) the same sequence of pulses
could be carried out and the wave packet would be adiabatically squeezed as it
moves along the light-induced channel into its final state.

The treatment presented here has been restricted to only three levels in the 
sodium dimer. There is always the possibility that other neighbouring energy 
levels could disturb the counterintuitive process by complicating the dressed 
state potential seen in Fig.~\ref{fig4}. This will be subject to further
investigation. Fortunately the scheme presented in this Letter is extremely
insensitive to the specific parameters (Rabi frequency and detuning) once the
appropriate regime has been discovered. Thus we expect that there is enough
freedom in the controlling parameters to avoid any detrimental effects from
other levels.

We have described a process for the efficient transfer of a wave packet from 
one molecular potential to another by means of light-induced potentials. We 
have demonstrated it with the sodium dimer using realistic pulse parameters and
potentials. A suitable name for the process is Adiabatic Passage by Light
Induced Potentials (APLIP). The process is not only efficient, but usable over a
wide range of counterintuitive pulse parameters. The range of parameters is
even wider than in a corresponding STIRAP excitation because we do not maintain
a precise two-photon resonance, and do not have the possibility of excitation
of neighbouring vibrational levels. The process can also be quite fast, almost
on vibrational timescales.

This work was supported by the United Kingdom Engineering and Physical
Sciences Research Council and the Academy of Finland. The authors wish to
thank Stig Stenholm and Nikolay Vitanov for helpful discussions.


\begin{references}
\bibitem{spont} We neglect spontaneous emission in this Letter, though it
can be included in the numerical treatment (see Ref.~\protect\cite{rpp}).
For short interaction times the influence of spontaneous emission is
small, partly because the intermediate state is not populated during the
process.
\end{references}
\end{document}